\documentclass[conference]{IEEEtran}
\usepackage[english]{babel}
\usepackage[utf8]{inputenc}
\usepackage[T1]{fontenc}
\usepackage{cite}
\usepackage{graphicx}
\usepackage{fixltx2e}

\begin{document}

\title{Impact of video quality and wireless network interface on power consumption of mobile devices}

\author{\IEEEauthorblockN{Norbert Zsak and Christian Wolff}
\IEEEauthorblockA{University of Regensburg, Universitätsstrasse 31, 93053 Regensburg, Germany\\
E-mail: \{norbert.zsak, christian.wolff\}@ur.de}}

\maketitle

\renewcommand\abstractname{Abstract}
\begin{abstract}
During the last years, many improvements were made to the hardware capability of mobile devices. As mobile software also became more interactive and data processing intensive, the increased power demand could not be compensated by the improvements on battery technology. Adaptive systems can help to balance the demand of applications with the limitations of battery resources. For effective systems, the influence of multimedia quality on power consumption of the components of mobile devices needs to be better understood. In this paper, we analyze the impact of video quality and wireless network type on the energy consumption of a mobile device. We have found that the additional power consumption is up to 38\% higher when a movie is played over a WiFi network instead from internal memory and 64\% higher in case of a mobile network (3G). We have also discovered that a higher movie quality not only affects the power consumption of the CPU but also the power consumption of the WiFi unit by up to 58\% and up to 72\% respectively on mobile networks.
\\\end{abstract}


\section{Introduction}
Since the use of smartphones and tablets became popular, many improvements have been made to the hardware of the mobile devices. Faster CPUs, new wireless interfaces and high resolution displays led to an increased power demand. In the same period of time, however, no comparably significant improvements of battery technology have been achieved. This leads to the problem that battery runtime steadily decreases as the demands of mobile software on hardware resources increase. This observation was made for multimedia applications or movie playback where complex computations have to be made or high amounts of data have to be transferred. In light of this development, the energy consumption of mobile devices has become an important area of needed development. 

Thus far, several approaches have been taken towards balancing users' requirements of high quality entertainment software with the limited hardware and battery resources of mobile systems. In many cases, these approaches focus on adaptive systems with an awareness of power consumption \cite{Ismail.2013}. To develop effective systems it is important to understand how the increased demand on multimedia applications and movie playback affect the power consumption of the particular components in a mobile device. 

While research has already shown that the codec and video quality of a movie leads to a severe increase in power demand of the computation components of a mobile device \cite{Sostaric.2010}, the impact of video codecs and quality to the other components, such as the wireless network interface have yet to be evaluated.

We therefore describe an experiment setup in which we evaluate the influence of both aspects, the video quality and the type of wireless network interface, on the power consumption of a mobile device.

\section{Experiment setup}
The tests were performed on a LG Nexus 5 smartphone. The high screen resolution of 1920x1080 pixel and the Quad-Core CPU \textit{Snapdragon 800} from \textit{Qualcomm} allow the device to play Full HD movies on the phone. To ensure stable energy consumption from the LCD, the brightness was fixed at 50\% for the experiment. The battery had a capacity of 2,300 mAh. In the following sections, we provide more detailed information about the test setup.

\subsection{Video setting}
The first three minutes from the computer animated movie \textit{Big Buck Bunny} \cite{BlenderFoundation} were taken as reference movie. The video stream was encoded in H.264 with a Full HD resolution, 25 fps and 9500 kbps bitrate. The audio stream was MPEG AAC encoded with 48 kHz sampling rate. The movie was transcoded into four various quality formats with the free software SUPER© \cite{eRightSoft}. The quality settings are described in table \ref{table:qualityformats}. As there is no video quality standard for mobile streaming, we followed the suggestions from YouTube \cite{Google}. The encoding of the audio stream remained unchanged.

\begin{table}
\caption{Tested video quality formats}
\label{table:qualityformats}
\centering
\begin{tabular}{|l|c|c|@{}}
\hline 
\textbf{Video Quality} & \textbf{Resolution} & \textbf{Bitrate (kbps)} \\ 
\hline 
\textbf{Low Definition (LD)} & 640x360 & 8000 \\ 
\hline 
\textbf{Standard Definition (SD)} & 854x480 & 5000 \\ 
\hline 
\textbf{HD ready (HD)} & 1280x720 & 2500 \\ 
\hline 
\textbf{Full HD (HD)} & 1920x1080 & 1000 \\ 
\hline 
\end{tabular}
\end{table}

\subsection{Wireless network interfaces}
To establish reference values for the power consumption of video playback without the use of wireless networks, we performed tests with playback from the internal memory. The WiFi test was performed using the 802.11n standard on the 5 GHz frequency. The smartphone was placed 3 meters away from the FRITZ!Box 7490 router. Thus, an average signal strength of --35 dBm was achieved. The mobile network test was performed on a HSPA (3G) network with a signal strength of --73 dBm.

\subsection{Power measurement}
The power consumption was measured with the app \textit{Trepn} from \textit{Qualcomm} \cite{QualcommTechnologies}. As the hardware platform of the LG Nexus 5 smartphone is also from Qualcomm, the software is able to read internal power management sensors for the calculations.
To prevent other programs on the smartphones from interfering with the measurements, all unneeded processes were closed before the test. The video stream was played on the app \textit{VLC}. For each combination of video quality and network mode, 12 runs were performed and the average power consumption was calculated.

\section{Results}
\begin{figure}[!t]
\centering
\includegraphics[width=3.4in]{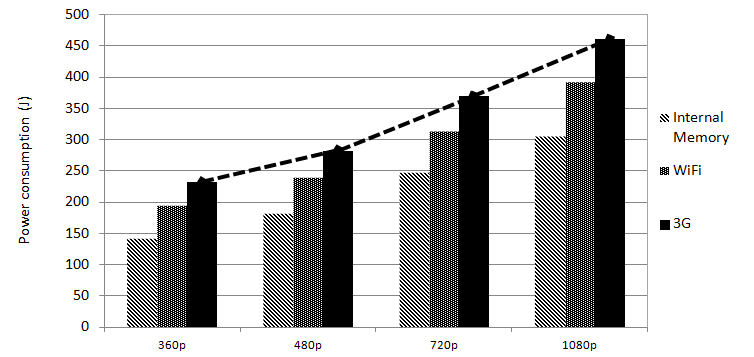}
\caption{Measurement results}
\label{fig:result}
\end{figure}

The results are listed in figure \ref{fig:result}. Though measuring the power consumption of a device with a software application may not be as precise as with special diagnostic hardware, the measurements provided reasonable results with V\textsubscript{x} < 0.05 for each video quality and network mode combination.

As expected, the power consumption of the smartphone depends heavily on the video quality. The power consumption for playing a 1080p movie is on average two times higher than for the 360p movie.
This relation also applies to movies played from the internal memory. While the 360p movie only consumed 783 mW on average, the 1080p movie required on average 1,696 mW. That additional consumption can be directly assigned to the CPU, GPU and memory because of the higher video decoding effort. All other components such as LCD, wireless network interfaces and audio were not affected by the higher video quality. 

We have analyzed the impact of video quality to the power consumption of the wireless mobile interface. The difference between a 3G and an internal memory playback of the 360p movie is 504 mW, while a difference of 865 mW could be observed with the 1080p movie between 3G and internal memory. The calculated difference reflects the net power consumption of the wireless network interface. The additional power consumption is on average up to 58\% higher on the WiFi and up to 72\% higher on the mobile network between the tested 360p and 1080p movies. This means that a higher video quality directly affects the power consumption of the wireless network interface.

We have also analyzed on how much the power consumption of a mobile device is increased when a movie is streamed over a mobile network or WiFi instead loaded from internal memory. The 1080p movie consumed on average 1,696 mW when played from the internal memory and 2,561 mW when streamed via a wireless network (3G).  This means that the streaming increased the overall power consumption of the device by 51\%. For the other video quality settings, the rate was between 49\% and 64\%. The use of a WiFi network for streaming video means an increase between 26\% and 38\% in overall power consumption compared to an internal memory playback. The use of a mobile network instead of WiFi for video streaming result in a considerable higher power consumption.

\section{Conclusion and limitations}
In this paper, we have not only shown that higher video quality scales with the power consumption of the CPU, GPU and memory of a mobile device, but also shown that the wireless network interface has a substantial impact on overall power consumption. In the field of mobile streaming, the requirements are highly dependent upon these factors. While in a mobile network we are being confronted with high power consumption and traffic limitations of the internet service provider, in a WiFi situation the two factors are less crucial. Therefore, adaptive systems are needed that decide which streaming parameters are best suited to a users' specific requirements and given situational capabilities.

Though the results of the experiment that were gathered by a software tool seem to be reasonable, they should be confirmed in a further study using hardware measurement tools. Additional studies should also evaluate the accuracy of software-based power measurement tools such as \textit{Trepn} or \textit{PowerTutor} \cite{Zhang} for various mobile devices. Additionally, the influence of other factors such as audio and video codec, type and signal strength of WiFi (e.g 802.11n, 802.11ac) and mobile networks (e.g UMTS, HSPA, LTE) and the hardware platform of the smartphone should also be evaluated.

\renewcommand\refname{References}
\bibliographystyle{./IEEEtran}
\bibliography{./IEEEabrv,./Literatur-ISPA2014}

\begin{thebibliography}{1}
\providecommand{\url}[1]{#1}
\csname url@samestyle\endcsname
\providecommand{\newblock}{\relax}
\providecommand{\bibinfo}[2]{#2}
\providecommand{\BIBentrySTDinterwordspacing}{\spaceskip=0pt\relax}
\providecommand{\BIBentryALTinterwordstretchfactor}{4}
\providecommand{\BIBentryALTinterwordspacing}{\spaceskip=\fontdimen2\font plus
\BIBentryALTinterwordstretchfactor\fontdimen3\font minus
  \fontdimen4\font\relax}
\providecommand{\BIBforeignlanguage}[2]{{%
\expandafter\ifx\csname l@#1\endcsname\relax
\typeout{** WARNING: IEEEtran.bst: No hyphenation pattern has been}%
\typeout{** loaded for the language `#1'. Using the pattern for}%
\typeout{** the default language instead.}%
\else
\language=\csname l@#1\endcsname
\fi
#2}}
\providecommand{\BIBdecl}{\relax}
\BIBdecl

\bibitem{Ismail.2013}
M.~N. Ismail, R.~Ibrahim, and {Md Fudzee, Mohd Farhan}, ``A survey on content
  adaptation systems towards energy consumption awareness,'' \emph{Advances in
  Multimedia}, vol. 2013, no.~3, pp. 1--8, 2013.

\bibitem{Sostaric.2010}
D.~Sostaric, D.~Vinko, and S.~Rimac-Drlje, ``Power consumption of video
  decoding on mobile devices,'' in \emph{ELMAR, 2010 PROCEEDINGS}, 2010, pp.
  81--84.

\bibitem{BlenderFoundation}
\BIBentryALTinterwordspacing
{Blender Foundation}, ``Big buck bunny.'' [Online]. Available:
  \url{\url{http://www.bigbuckbunny.org/}}
\BIBentrySTDinterwordspacing

\bibitem{eRightSoft}
\BIBentryALTinterwordspacing
eRightSoft, ``Super{\copyright}.'' [Online]. Available:
  \url{\url{http://www.erightsoft.com/SUPER.html}}
\BIBentrySTDinterwordspacing

\bibitem{Google}
\BIBentryALTinterwordspacing
Google, ``Advanced encoding settings: Recommended bitrates, codecs, and
  resolutions, and more.'' [Online]. Available:
  \url{\url{https://support.google.com/youtube/answer/1722171?hl=en}}
\BIBentrySTDinterwordspacing

\bibitem{QualcommTechnologies}
\BIBentryALTinterwordspacing
{Qualcomm Technologies}, ``Trepn profiler.'' [Online]. Available:
  \url{\url{https://developer.qualcomm.com/mobile-development/increase-app-performance/trepn-profiler}}
\BIBentrySTDinterwordspacing

\bibitem{Zhang}
L.~Zhang, B.~Tiwana, Z.~Qian, Z.~Wang, R.~P. Dick, Z.~M. Mao, and L.~Yang,
  ``Accurate online power estimation and automatic battery behavior based power
  model generation for smartphones,'' in \emph{the eighth IEEE/ACM/IFIP
  international conference}, T.~Givargis and A.~Donlin, Eds., p. 105.

\end{thebibliography}

\end{document}